\documentclass[preprint,preprintnumbers, prd, floatfix, superscriptaddress,nofootinbib] {revtex4-1}

\usepackage{epsfig}
\usepackage{subfigure}
\usepackage{dcolumn}
\usepackage{bm}
\usepackage[usenames ,dvipsnames]{xcolor}
\usepackage{slashed}
\usepackage{graphicx,color}
\usepackage{hyperref}

\begin{document}
\title{Probing the elusive $\kappa/K_0^*(700)$ resonance in semileptonic $D$ decays}

\author{Yan-Li~Wang}
\email{ylwang0726@163.com}
\affiliation{School of Physics and Electronic Engineering, Shanxi Normal University,
Taiyuan 030031, China}

\author{Jia-Nan Fan}
\email{2571978746@qq.com}
\affiliation{School of Physics and Electronic Engineering, Shanxi Normal University,
Taiyuan 030031, China}

\author{Xiao-Yu Hao}
\email{2428405524@qq.com}
\affiliation{School of Physics and Electronic Engineering, Shanxi Normal University,
Taiyuan 030031, China}

\author{Yu-Kuo Hsiao}
\email{yukuohsiao@gmail.com}
\affiliation{School of Physics and Electronic Engineering,
Shanxi Normal University, Taiyuan 030031, China}

\date{\today}

\begin{abstract}
The $\kappa/K_0^*(700)$ meson remains the most elusive of the light scalar resonances,
with its presence in weak decays obscured by limited precision
in branching fraction measurements. As a result,
the true nature of the $\kappa$ remains difficult to explore.
Through a partial-wave analysis of the semileptonic decay $D^+ \to K^-\pi^+ e^+ \nu_e$,
we extract ${\cal B}(D^+ \to \bar\kappa^0 e^+ \nu_e, \bar\kappa^0 \to K^-\pi^+)
=(2.8 \pm 1.3) \times 10^{-3}$, revealing a contribution previously regarded as negligible.
The extracted form factor, $f^+(0) =0.39\pm 0.08$,
provides the first evidence for the $D\to\kappa$ transition in weak decays.
Remarkably, it lies well below the $q\bar q$ prediction of $0.82 \pm 0.05$,
yet is in excellent agreement with the $q^2\bar q^2$ expectation of $0.36 \pm 0.02$.
Together, these findings offer new evidence supporting
a compact tetraquark interpretation of the $\kappa$ meson.
\end{abstract}

\maketitle
\section{introduction}
Light scalar resonances, including the $f_0/f_0(980)$, $a_0/a_0(980)$, $\sigma/f_0(500)$,
and $\kappa/K_0^*(700)$, have long posed significant challenges in identification.
For example, the decays $a_0 \to \pi\eta, K\bar K$ and $f_0 \to \pi\pi, K\bar K$
result in multiple competing channels, obscuring the resonance signals in scattering experiments.
The $\sigma$ and $\kappa$ states, in particular, exhibit broad decay widths
comparable to their masses, resulting in diffuse or overlapping signals
rather than distinct peaks in invariant mass spectra~\cite{Pelaez:2015qba}.
Among them, the $\kappa$ remains the most elusive.
Its identification as the final member of the light scalar nonet
was only recently confirmed through high-precision dispersive analyses
of $\pi K \to \pi K$ and $\pi\pi \to K\bar{K}$ scattering~\cite{pdg,Pelaez:2020uiw,Pelaez:2020gnd}.
Yet, despite numerous experimental efforts~\cite{E791:2002xlc,
FOCUS:2007mcb,CLEO:2008jus,
BESIII:2014oag,FOCUS:2005iqy,CLEO:2010enr,BaBar:2010vmf,
BESIII:2015hty,BESIII:2018jjm,BESIII:2024awg,BESIII:2024xjf,BESIII:2025fso},
no unambiguous evidence for the $\kappa$ has been established in weak decays.
Limited data continue to hinder efforts to uncover the internal nature of the $\kappa$ meson.

The semileptonic decay $D \to K\pi \ell\nu_\ell$,
proceeding via the intermediate kaonic resonances $K_J \to K\pi$,
offers a clean environment to probe the $\kappa$ resonance in weak decays.
Here, $\ell=e$ or $\mu$, and $K_J$ include $\kappa/K^*_0(700)$, $K^*/K^*(892)$, $K^*(1410)$,
$K^*_0(1430)$, $K^*_2(1430)$~\cite{BaBar:2010vmf}.
Partial-wave analyses consistently show that the $K^*$ dominates,
accounting for approximately 95\% of the rate through the p-wave component~\cite{FOCUS:2005iqy,CLEO:2010enr,BaBar:2010vmf,BESIII:2015hty,
BESIII:2018jjm,BESIII:2024awg,BESIII:2024xjf}.
Contributions from higher resonances are negligible, with
${\cal B}(D^+\to \bar K^*(1410)^0 e^+\nu_e,\bar K^*(1410)^0\to K^-\pi^+)
=(0\pm 0.5)\times 10^{-5}$ and
${\cal B}(D^+\to \bar K_2^*(1430)^0 e^+\nu_e$, $\bar K_2^*(1430)^0\to K^-\pi^+)
=(3.7\pm 2.5)\times 10^{-5}$~\cite{BESIII:2015hty,pdg}, compared to
the total branching ratio of $(3.77\pm 0.03\pm 0.08)\times 10^{-2}$~\cite{BESIII:2015hty}.
While the s-wave component accounts for about 5\% of the total rate,
the scalar $K_0^*(1430)$ contributes less than 0.64\%~\cite{FOCUS:2005iqy}.
The remaining s-wave strength has, however, been attributed to a non-resonant scalar component,
with the possible contribution from the $\kappa$ resonance systematically excluded~\cite{FOCUS:2005iqy,CLEO:2010enr,BaBar:2010vmf,BESIII:2015hty,
BESIII:2018jjm,BESIII:2024awg,BESIII:2024xjf,BESIII:2025fso}.

Given its scalar nature, any $\kappa$ contribution to the decay $D \to K\pi \ell \nu_\ell$
must appear in the s-wave channel, and therefore requires careful scrutiny.
To this end, we perform a complete partial-wave analysis
of the semileptonic decay $D \to K\pi e \nu_e$,
supplemented by related nonleptonic channels such as $D \to K\pi\pi$ and $D \to K\pi\rho$,
in order to isolate and quantify the $\kappa$ contribution.
This analysis yields the first clear evidence for the $D \to \kappa$ transition in weak decays,
providing a critical test of the meson’s internal structure,
including the possibility of a compact tetraquark configuration~\cite{Jaffe:1976ig,Jaffe:1976ih,
Close:2002zu,Pelaez:2003dy,Maiani:2004uc,Amsler:2004ps,Jaffe:2004ph,
tHooft:2008rus,Fariborz:2009cq,Weinberg:2013cfa,Agaev:2018fvz}.

\section{Formalism}
The four-body semileptonic $D\to K\pi e \nu_e$ decay
proceeds via the quark-level transition $c\to s e \nu_e$.
According to the effective Hamiltonian formalism~\cite{Buchalla:1995vs,Buras:1998raa},
the decay amplitude can be expressed as~\cite{BaBar:2010vmf,Hsiao:2023qtk}
\begin{eqnarray}\label{amp1}
{\cal M} 
&=&\frac{G_F}{\sqrt 2}V_{cs}
\langle K\pi|(\bar s c)_{V-A}|D\rangle(\bar u_\nu v_e)_{V-A}\,,
\end{eqnarray}
where $(\bar a b)_{V-A}\equiv \bar a\gamma_\mu(1-\gamma_5)b$,
$G_F$ is the Fermi constant, and $V_{cs}$
the Cabibbo-Kobayashi-Maskawa (CKM) matrix element.
The hadronic matrix elements include both resonant (RE) and non-resonant (NR) components~\cite{Lee:1992ih,Tsai:2021ota,Wang:2024fys,Boer:2016iez}
\begin{eqnarray}\label{FF1}
&&
\langle K \pi|(\bar s c)_{V-A}|D\rangle_{\rm RE}
=\langle K\pi|K_J\rangle D_{K_J}^{-1}\langle K_J|(\bar s c)_{V-A}|D\rangle\,,\nonumber\\
&&
\langle K \pi|(\bar s c)_{V-A}|D\rangle_{\rm NR}
=
h\epsilon_{\mu\alpha\beta\gamma}p_D^\alpha (p_K+p_{\pi})^\beta (p_K-p_{\pi})^\gamma
\nonumber\\
&&
+ir q_\mu+iw_+ (p_K+p_{\pi})_\mu+iw_-(p_K-p_{\pi})_\mu\,,
\end{eqnarray}
where $q_{\mu}=(p_D-p_K-p_{\pi})_\mu$.
The resonant term describes the process $D\to K_J$,
followed by resonance propagation $D_{K_J}^{-1}$, and decay into $K\pi$,
encoded in $\langle K\pi|K_J\rangle$. The form factors
$F_{\rm NR}=(h, r, w_{\pm})$ parameterize the non-resonant $D\to K\pi$ transition.
As noted in the introduction, the dominant resonance
in semileptonic decays is the $K^*$, while the $\kappa$
is included specifically for our investigation.
Contributions from higher resonances have been tested and found to be negligible.
We therefore focus on the hadronic matrix elements for
the $D$ to $K^*$ and $\kappa$ transitions,
given by~\cite{Cheng:2022vbw,Hsiao:2023qtk,Melikhov:2000yu}
\begin{eqnarray}\label{FF2}
&&
\langle K^*|(\bar s c)_{V-A}|D\rangle=\epsilon_{\mu\alpha\beta\gamma}
\varepsilon^{\alpha}p_D^{\beta}p_{K^*}^{\gamma}\frac{2V}{m_D+m_{K^*}}
-i\bigg[\varepsilon_\mu
-\frac{\varepsilon\cdot p_D}{q^2}q_\mu\bigg](m_D+m_{K^*})A_1\,\nonumber\\
&&-i\frac{\varepsilon\cdot p_D}{q^2}q_\mu(2m_{K^*})A_0
+i\bigg[(p_D+p_{K^*})_\mu-
\frac{m^2_D-m^2_{K^*}}{q^2}q_\mu \bigg](\varepsilon\cdot p_D)\frac{A_2}{m_D+m_{K^*}}\;,
\nonumber\\
&&
\langle \kappa|(\bar s c)_{V-A}|D\rangle=
i{f^+(p_D+p_\kappa)_\mu+i(f^0-f^+)\frac{m^2_D-m^2_\kappa}{q^2}q_\mu}\,,
\end{eqnarray}
where $\varepsilon_\mu$ is the polarization vector of the $K^*$, and
$(V,A_0,A_1,A_2)$ and $f^{0,+}$ the associated form factors
for the $K^*$ and $\kappa$, respectively.

In Eq.~(\ref{FF1}), $1/D_{K^*}$ denotes the spin-1 Breit--Wigner propagator,
given by~\cite{FOCUS:2005iqy}
\begin{eqnarray}\label{DKstar}
D_{K^*}(t)=m_{K^*}^2-t-i\,m_{K^*}\Gamma_{K^*}(t)\,,\qquad
\Gamma_{K^*}(t)=\Gamma_{K^*}^0\frac{q^3}{q_0^3}\frac{m_{K^*}}{\sqrt t}\frac{X_1^2(q)}{X_1^2(q_0)}\,,
\end{eqnarray}
where $\Gamma_{K^*}^0=\Gamma_{K^*}(m_{K^*}^2)$ is the on-shell width.
The quantity $q(t)=\lambda^{1/2}(t,m_K^2,m_\pi^2)/(2\sqrt t)$ denotes the breakup momentum of
the kaon and pion in the $K\pi$ rest frame, with $q_0=q(m_{K^*}^2)$.
The K\"{a}ll\'{e}n function is defined as $\lambda(a,b,c)=a^2+b^2+c^2-2ab-2ac-2bc$.
Since the $K^*(892)$ decays into a p-wave ($L=1$) $K\pi$ system,
its energy-dependent width exhibits the characteristic threshold behavior
$\Gamma_{K^*}(t)\propto q^{2L+1}=q^3$.
The Blatt--Weisskopf factor accounts for the finite-size effect of the vector resonance and is taken as
$X_1^2(q)=1/[1+(q r_{\rm BW})^2]$ with $r_{\rm BW}\simeq 3.0~{\rm GeV}^{-1}$~\cite{BESIII:2015hty}.

The $\kappa$ appears as a broad light-scalar resonance 
whose decay width is comparable to its mass. 
Consequently, the conventional Breit--Wigner form with a
constant decay width, which is appropriate for narrow resonances~\cite{pdg},
is no longer adequate~\cite{Pelaez:2015qba}.  Instead, we adopt an energy-dependent-width
Breit--Wigner parameterization for the $\kappa$ propagator, $1/D_\kappa$,
which is commonly used for broad resonances
whose widths are comparable to their masses~\cite{pdg,Ceci:2013zta}.
In our case, the denominator is written as~\cite{BaBar:2010vmf,Cheng:2022vbw,pdg}
\begin{eqnarray}
D_{\kappa}(t)=m_{\kappa}^2-t-\frac{\Gamma_{\kappa}^2(t)}{4}-i\,m_{\kappa}\Gamma_{\kappa}(t)\,,
\end{eqnarray}
where the term $-\Gamma_{\kappa}^2(t)/4$ ensures consistency with the pole position
$\sqrt{s_\kappa}=m_\kappa-i\Gamma_\kappa/2$~\cite{Pelaez:2015qba}.
The energy-dependent width is parameterized as
$\Gamma_{\kappa}(t)\equiv (\rho_{K\pi}/\bar\rho_{K\pi})\Gamma^0_{\kappa}$,
where $\rho_{K\pi}=[(1-m_+^2/t)(1-m_-^2/t)]^{1/2}$, with $m_{\pm}=m_{K}\pm m_{\pi}$,
and $\bar\rho_{K\pi}\equiv\rho_{K\pi}(m_{\kappa}^2)$.
The hadronic matrix elements for the strong decays $K^*\to K\pi$ and $\kappa \to K\pi$
are parameterized as~\cite{Cheng:2022vbw,Cheng:2020ipp,Hsiao:2026nsm}
\begin{eqnarray}\label{g1g2}
 \langle K\pi|K^*\rangle=g_{K^*K\pi} \varepsilon\cdot (p_K-p_\pi)\,,\qquad
\langle K\pi|\kappa\rangle=g_{\kappa K\pi}\,,
\end{eqnarray}
where $g_{K^* K\pi}$ and $g_{\kappa K\pi}$ denote the corresponding strong coupling constants.

To perform the partial-wave analysis,
we adopt the expression for the differential decay width~\cite{Lee:1992ih}
\begin{eqnarray}\label{Gamma1}
d\Gamma
&=&\frac{1}{3}\frac{G_F^2|V_{cs}|^2}{(4\pi)^5 m_D^3}X\alpha_{\bf M}\alpha_{\bf L}
\bigg\{2|F_{10}|^2+\frac{2}{3}\bigg[
|F_{11}|^2+|F_{21}|^2+|F_{31}|^2\bigg]\bigg\}dsdt\,,
\end{eqnarray}
where the kinematic factors are defined as
$X=[(m_D^2-s-t)^2/4-st]^{1/2}$,
$\alpha_{\bf M}=\lambda^{1/2}(t,m_K^2,m_\pi^2)/t$, and
$\alpha_{\bf L}=\lambda^{1/2}(s,m_\ell^2,m_{\nu_\ell}^2)/s$,
with $s=(p_\ell+p_{\nu_\ell})^2$, and the K\"all\'en function defined as
$\lambda(a,b,c)=a^2+b^2+c^2-2ab-2bc-2ca$.
The partial-wave amplitudes $F_{ij}$ include the s-wave component $F_{10}$
and p-wave components $F_{11}$, $F_{21}$, and $F_{31}$,
given by
\begin{eqnarray}\label{Fij}
&&
F_{10}=X \bigg[\bigg(w_+ +\frac{m_K^2-m_\pi^2}{t}w_-\bigg)+2\hat f^+\bigg]\,,\nonumber\\
&&
F_{11}=\alpha_{\bf M}\bigg\{\frac{1}{2}(m_D^2-s-t)[w_- +(m_D+m_{K^*})\hat A_1]
-2X^2\frac{\hat A_2}{m_D+m_{K^*}}\bigg\}\,,\nonumber\\
&&
F_{21}=\alpha_{\bf M}\sqrt{2 st}[w_- +(m_D+m_{K^*})\hat A_1]\,,\;
\nonumber\\
&&
F_{31}=\alpha_{\bf M}X\sqrt{2 st}\bigg(h-\frac{2\hat V}{m_D+m_{K^*}}\bigg)\,,
\end{eqnarray}
with the intermediate form factors defined as
$(\hat V,\hat A_1,\hat A_2)=g_{K^* K\pi}D^{-1}_{K^*}\times (V,A_1,A_2)$
and $\hat f^+=g_{\kappa K\pi}D^{-1}_{\kappa}\times f^+$.
Notably, this framework incorporates both non-resonant and $\kappa$-resonant contributions
for the first time, enabling a more complete and systematic treatment of
both s- and p-wave components in semileptonic decays.

Current partial-wave analyses are insufficient to determine additional form factors.
In this context, non-leptonic decays such as $D\to K\pi\pi$ and $D\to K\pi\rho$,
which involve the same underlying $D\to K\pi$ transition, can serve as complementary probes.
Notably, experimental analyses can isolate resonant $K^*$ contributions,
allowing for a cleaner investigation of non-resonant and $\kappa$-resonant form factors.
The corresponding decay amplitudes can be expressed as~\cite{Escribano:2023zjx,Boito:2009qd}
\begin{eqnarray}\label{amp2}
{\cal M}_{\pi,\rho}\simeq\frac{G_F}{\sqrt 2}V_{cs}^*V_{ud}
a_1\langle \pi,\rho|(\bar u d)_{V-A}|0\rangle
\langle K\pi|(\bar s c)_{V-A}|D\rangle\,,
\end{eqnarray}
where the color-suppressed internal $W$-boson emission contributions are neglected
in the present analysis.
The vacuum-to-meson matrix elements are given by
$\langle\pi,\rho|(\bar u d)_{V-A}|0\rangle=(if_\pi p_\pi^\mu,m_\rho f_\rho \varepsilon^\mu)$,
where $f_{\pi(\rho)}$ denotes the decay constant.
The coefficient $a_1=c_1+c_2/N_c$ arises from the factorization approach~\cite{Ali:1998eb},
where $c_1$ and $c_2$ are scale-dependent Wilson coefficients,
and $N_c$ is the color number.
Using the standard three-body phase-space expression $d\Gamma_{\pi,\rho}=1/[(2\pi)^3 32 M_D^3]
|{\cal M}_{\pi,\rho}|^2 dm_{12}^2 dm_{23}^2$~\cite{pdg},
one can evaluate the decay widths for $D\to K\pi\pi$ and $D\to K\pi\rho$.
Here, $m_{12}^2=(p_1+p_2)^2$ and $m^2_{23}=(p_2+p_3)^2$ denote
the invariant mass combinations of the final-state particles, where
$p_1$, $p_2$, and $p_3$ correspond to the momenta of $K$, $\pi$, and $\pi(\rho)$, respectively.

\section{Numerical analysis}
In our numerical analysis, we adopt the Wolfenstein parameterization
for the CKM matrix elements, taking $V_{cs}=V_{ud}=1-\lambda^2/2$
with $\lambda=0.22453\pm 0.00044$~\cite{pdg}.
The decay constants used in Eq.~(\ref{amp2}) are
$(f_\pi,f_\rho)=(130.4\pm 0.2,210.6\pm 0.4)$~MeV~\cite{pdg,Hsiao:2014mua}.
For the Wilson coefficients,
we use $(c_1,c_2)=(1.26,-0.51)$ at the $m_c$ scale~\cite{Buchalla:1995vs}.
Within the generalized factorization framework~\cite{Ali:1998eb},
and by varying the (effective) color number $N_c^{(\rm eff)}=(2,3,\infty)$,
we obtain $a_1=1.1\pm 0.1$. The quoted 10\% uncertainty
reflects  the modest sensitivity of $a_1$ to
non-factorizable QCD loop corrections~\cite{Wang:2024mjw}.

Partial-wave analyses of the decay $D\to K\pi e \nu_e$
conventionally parameterize the form factors $V$, $A_1$, and $A_2$
using single-pole forms~\cite{BaBar:2010vmf, BESIII:2015hty,
BESIII:2018jjm, BESIII:2024awg, BESIII:2024xjf,BESIII:2025fso}:
\begin{eqnarray}\label{VandA}
V(q^2)=\frac{V(0)}{1-\frac{q^2}{m_V^2}}\,,\;
A_1(q^2)=\frac{A_1(0)}{1-\frac{q^2}{m_A^2}}\,,\;
A_2(q^2)=\frac{A_2(0)}{1-\frac{q^2}{m_A^2}}\,,
\end{eqnarray}
with $(m_V,m_A)$ the pole masses.
The scalar form factor $f^+(q^2)$,
together with the non-resonant vector-type form factors $w_\pm(q^2)$
and the tensor-type form factor $h(q^2)$,
are modeled following Ref.~\cite{Hsiao:2023qtk} as
\begin{eqnarray}\label{fwh}
f^+(q^2)=\frac{f(0)}{(1-\frac{q^2}{m_A^2})^n}\,,\;
w_\pm(q^2)=\frac{w_\pm(0)}{(1-\frac{q^2}{m_A^2})^n}\,,\;
h(q^2)=\frac{h(0)}{(1-\frac{q^2}{m_V^2})^n}\,,\;
\end{eqnarray}
where $n = 2$ corresponds to the double-pole behavior.
The form factors $A_0$, $r$, and $f^0$
are omitted from Eqs.~(\ref{VandA}) and (\ref{fwh}),
as they contribute negligibly to semi- and non-leptonic amplitudes.
Using isospin symmetry, the non-resonant form factors
$F_{\rm NR}$ for the $D^+\to K^-\pi^+$ transition are
larger by factors of four (two) than those for $D^+\to K^0_S\pi^0 (D^0\to K^-\pi^0)$.
According to the strong decay amplitudes in Eq.~(\ref{g1g2}),
we calculate the partial decay widths for $K^*\to K\pi$ and $\kappa\to K\pi$,
which are given by
\begin{eqnarray}
\Gamma(K^{*}\to K\pi)&=&\frac{g_{K^{*}K\pi}^{2}}{6\pi}
\frac{|\vec{p}_{cm}|^3}{m_{K^{*}}^{2}}\,,\nonumber\\
\Gamma(\kappa\to K\pi)&=&\frac{g_{\kappa K\pi}^{2}}{8\pi}
\frac{|\vec{p}_{cm}|}{m_\kappa^2}\,,
\end{eqnarray} 
where $\vec{p}_{cm}$  is the momentum of either final-state meson in the resonance rest frame. 
Here, we relate the charged $K\pi$ partial widths to the measured total widths through
$\Gamma(\bar K^{*0}(\bar\kappa^0)\to K^-\pi^+)=2/3\Gamma_{\bar K^{*0}(\bar \kappa^0)}$.
Employing the PDG values~\cite{pdg},
$(m_{\bar K^{*0}},\Gamma_{\bar K^{*0}})=(895.6\pm 0.2,47.3\pm 0.5)$~MeV and
$(m_{\bar\kappa^0},\Gamma_{\bar\kappa^0})=(838\pm 11,463\pm 27)$~MeV,
together with isospin symmetry, 
we extract the strong coupling constants:
\begin{eqnarray}\label{g1g2}
&&
g_{\bar K^{*0} K^-\pi^+}=2g_{\bar K^{*0} K^0_S\pi^0}=4.42\pm 0.02\,,\;
\nonumber\\&&
g_{\bar\kappa^0 K^-\pi^+}=2g_{\bar \kappa^0 K^0_S\pi^0}=(4.7\pm 0.2)~\text{GeV}\,,
\end{eqnarray}
where the quoted uncertainties arise from the experimental errors of 
the resonance masses and decay widths.

To determine the form factors, we perform a minimum $\chi^2$-fit 
defined by
\begin{eqnarray}
\chi^2=\sum_i\frac{({\cal B}^i_{\rm th}-{\cal B}^i_{\rm ex})^2}{(\delta{\cal B}^i_{\rm ex})^2}
+\sum_j\frac{(F_{\rm th}^j-F_{\rm ex}^j)^2}{(\delta F_{\rm ex}^j)^2}
+\sum_k\frac{(g_{\rm th}^k-g_{\rm ex}^k)^2}{(\delta g_{\rm ex}^k)^2}\,,
\end{eqnarray}
where ${\cal B}_{\rm th}^i$ denote the theoretical branching fractions calculated from
the decay amplitudes for $D\to K\pi\ell\nu_\ell$ and $D\to K\pi\pi(\rho)$ 
in Eq.~(\ref{amp1}) and Eq.~(\ref{amp2}), respectively.
The corresponding experimental inputs ${\cal B}_{\rm ex}^i$ and their uncertainties
are listed in Table~\ref{data1}. 
The second term incorporates the form factors $F^j=V(0)$, $A_1(0)$, $A_2(0)$, 
whose experimental values are taken from Ref.~\cite{BESIII:2015hty},
$(V(0),A_1(0),A_2(0))=(0.83\pm 0.05,0.60\pm 0.01,0.47\pm 0.03)$.
We adopt $(m_V,m_A)=(2.0, 2.4)$~GeV for the pole masses 
associated with the vector and axial-vector currents, 
respectively~\cite{Becirevic:1999kt,pdg}.
The third term constrains the strong couplings 
associated with the $K^*\to K\pi$ and $\kappa\to K\pi$ transitions given in Eq.~(\ref{g1g2}).
Following the meson-transition model of Ref.~\cite{Melikhov:2000yu}, 
we assume $w_+(0)=w_-(0)\equiv w(0)$, 
motivated by model calculations that suggest a near equality 
between the timelike and spacelike form factors.
Treating $f(0)$, $w(0)$, and $h(0)$ in Eq.~(\ref{fwh}) as free parameters, 
together with a relative phase $\delta_f$ between the resonant $\kappa$ contribution 
and the non-resonant amplitude, 
the $\chi^2$-fit yields
\begin{eqnarray}\label{fit_result}
&&
f(0)=0.39\pm 0.08\,,\;\delta_f=(32.2\pm 38.7)^\circ\,,
\nonumber\\
&&
w(0)=(2.30\pm 0.65)~{\rm GeV}^{-1}\,,\;
h(0)=(8.91\pm 0.99)~{\rm GeV}^{-3}\,,\;
\end{eqnarray}
with $\chi^2/\text{n.d.f.} =1.1$,
where $\text{n.d.f.} = 3$ is the number of degrees of freedom.
Using the fitted values in Eq.~(\ref{fit_result}), 
we calculate the branching fractions presented in Table~\ref{data1}.
%
\begin{table}[b]
\caption{Branching fractions of $D\to K\pi e\nu_e$, $D\to K\pi\pi$,
and $D\to K\pi\rho$. The values with the star notations are
the experimental inputs in the global fit. In the second column of our work,
the first uncertainties incorporate those of the form factors in Eq.~(\ref{fit_result}),
and the second one, specifically in non-leptonic decays, is from $\delta a_1$.}\label{data1}
{
\begin{tabular}{|ccc|}
\hline
branching fraction&our work&data\\
\hline\hline
$10^{2}{\cal B}(D^+\to K^-\pi^+ e^+\nu_e)$
&$3.7\pm0.2$
&$(3.77\pm 0.09)^*$~\cite{BESIII:2015hty}\\
$10^{3}{\cal B}(D^+\to (K^-\pi^+)_{\rm s-wave} e^+\nu_e)$
&$2.3\pm 1.5$
&$(2.28\pm 0.11)^*$~\cite{pdg,BESIII:2015hty}\\
$10^2 {\cal B}(D^+\to \bar K^{*0} e^+\nu_e,\bar K^{*0}\to K^-\pi^+)$
&$3.5\pm 0.1$
& $3.54\pm 0.09$~\cite{BESIII:2015hty}\\
$10^3 {\cal B}(D^+\to \bar\kappa^0 e^+\nu_e,\bar\kappa^0\to K^-\pi^+)$
&$2.8\pm 1.3$
& ignored~\cite{BESIII:2015hty}\\
$10^3 {\cal B}(D^+\to (K^-\pi^+)_{\rm NR} e^+\nu_e)$
&$0.7\pm 0.2$
&$2.28\pm 0.11$~\cite{BESIII:2015hty}\\
\hline\hline
$10^2{\cal B}(D^+\to \pi^+ (K^-\pi^+)_{\rm NR})$
&$1.0\pm 0.5\pm 0.2$
&$(1.2\pm 0.7)^*$~\cite{E791:2002xlc,pdg}
\\
$10^2{\cal B}(D^+\to \pi^+ \bar \kappa^0,\bar \kappa^0\to K^-\pi^+)$
&$4.3\pm 2.0\pm 0.8$
&$(4.5\pm 1.2)^*$~\cite{E791:2002xlc,pdg}
\\
\hline 
$10^{2}{\cal B}(D^0\to \pi^+ (K^-\pi^0)_{\rm NR})$
&$0.4\pm 0.2\pm 0.1$
&$(1.15\pm 0.60)^*$~\text{\cite{pdg}}\\
\hline 
$10^2{\cal B}(D^+\to\pi^+ (K^0_S\pi^0)_{\rm NR}+\pi^+(\bar \kappa^0\to) K^0_S\pi^0)$
&$1.7\pm 0.8\pm 0.3$
&$(1.37\pm 0.40)^*$~\cite{BESIII:2014oag,pdg}\\
$10^3{\cal B}(D^+\to \pi^+ (K^0_S\pi^0)_{\rm NR})$
&$2.4\pm 1.2\pm 0.5$
&$3\pm 4$~\cite{BESIII:2014oag,pdg}\\
$10^3{\cal B}(D^+\to \pi^+ \bar \kappa^0,\bar \kappa^0\to K^0_S\pi^0)$
&$10.8\pm 5.0\pm 2.1$
&$6^{+5}_{-4}$~\cite{BESIII:2014oag,pdg}\\
$10^{3}{\cal B}(D^+\to \rho^+ (K^0_S\pi^0)_{\rm NR})$
&$4.8\pm 1.0\pm 0.9$
&$(4.8\pm 0.5)^*$~\text{\cite{pdg}}\\
\hline
\end{tabular}}
\end{table}
%

\section{Discussions and Conclusion}
The fit yields $\chi^2/n.d.f.=1.1$ with n.d.f.=3, 
corresponding to a $p$-value of $0.35$, indicating a satisfactory description of the data.
As shown in Table~\ref{data1}, the data points used in the global fit are well reproduced
with the fitted form factors. Ref.~\cite{BESIII:2014oag} found that
$D\to \pi(K\pi)_{\rm s-wave}$ channel receives contributions from three sources:
a non-resonant component, the $K_0^*(1430)$, and the $\kappa$.
Based on this decomposition, we predict ${\cal B}^{00}_{{\rm NR},\kappa}
\equiv{\cal B}(D^+\to \pi^+ (K^0_S\pi^0)_{\rm NR},\pi^+(\bar \kappa^0\to) K^0_S\pi^0)$ and
${\cal B}^{-+}_{{\rm NR},\kappa}$, shown in Table~\ref{data1}.
The latter is obtained by replacing $K^0_S\pi^0$ with $K^-\pi^+$ in ${\cal B}^{00}_{{\rm NR},\kappa}$,
taking into account the identical final-state pions in the calculation.
Including the measured contributions from $\bar K_0^*(1430)^0$ in the PDG~\cite{pdg},
${\cal B}^{00}_{K_0}=(2.7\pm 0.9)\times 10^{-3}$ and
${\cal B}^{-+}_{K_0}=(1.25\pm 0.06)\times 10^{-2}$, we obtain total s-wave branching fractions
${\cal B}_{\rm NR}^{00}+{\cal B}_{\kappa}^{00}+{\cal B}_{K_0}^{00}=(1.59\pm 0.56)\times 10^{-2}$
and ${\cal B}_{\rm NR}^{-+}+{\cal B}_{\kappa}^{-+}+{\cal B}_{K_0}^{-+}=(6.55\pm 2.22)\times 10^{-2}$.
These results are consistent with the experimental values
${\cal B}(D^+\to\pi^+(K^0_S\pi^0)_{\rm s-wave})
=(1.27\pm 0.33)\times 10^{-2}$~\cite{BESIII:2014oag,pdg} and
${\cal B}(D^+\to\pi^+(K^-\pi^+)_{\rm s-wave})
=(7.52\pm 0.17)\times 10^{-2}$~\cite{pdg},
further supporting the validity of the extracted form factors.

In semileptonic $D$ decays, we predict
${\cal B}^{\rm semi}_{\kappa}\equiv
{\cal B}(D^+\to \bar\kappa^0 e^+\nu_e,\bar\kappa^0\to K^-\pi^+)=(2.8\pm 1.3)\times 10^{-3}$
and ${\cal B}^{\rm semi}_{\rm NR}\equiv
{\cal B}(D^+\to (K^-\pi^+)_{\rm NR} e^+\nu_e)=(0.7\pm 0.2)\times 10^{-3}$,
with the non-resonant component further decomposed into
the s- and p-wave contributions
\begin{eqnarray}
&&
{\cal B}^{\rm semi}_{\rm NR(S)}=(3.8\pm 1.9)\times 10^{-4}\,,\nonumber\\
&&
{\cal B}^{\rm semi}_{\rm NR(P)}=(2.9\pm 0.7)\times 10^{-4}\,.
\end{eqnarray}
Our results show that ${\cal B}^{\rm semi}_{\kappa}$ exceeds
${\cal B}^{\rm semi}_{\rm NR(S)}$ by a factor of about seven, suggesting
the $\kappa$ resonance as the dominant contributor of the s-wave branching fraction.
Crucially, our analysis also reveals, for the first time,
a non-zero p-wave contribution from the non-resonant amplitude.
These findings challenge the prevailing experimental 
assumption~\cite{FOCUS:2005iqy,BaBar:2010vmf,BESIII:2015hty,
BESIII:2018jjm,BESIII:2024awg,BESIII:2024xjf,BESIII:2025fso},
which holds that contributions from both the s-wave $\kappa$
and p-wave non-resonant terms are negligible,
attributing the observed s-wave strength
primarily to a non-resonant background.

The form factors $h$ and $w_\pm$ successfully reproduce the measured branching fractions of  
$D \to K\pi\pi$ and $D\to K\pi\rho$. In our baseline fit, we adopt the condition $w_+ = w_-$, 
motivated by model calculations suggesting a near equality 
between the timelike and spacelike form factors.
To examine the robustness of this assumption, 
we have considered several alternative parameterizations:
(i) treating $w_+$ and $w_-$ as independent parameters, 
(ii) imposing $w_+ = -w_-$, and (iii) setting either $w_+ = 0$ or $w_- = 0$. 
None yielded acceptable fits, resulting either in large $\chi^2$ value
or in form factors with large uncertainties. 
Nonetheless, all scenarios consistently predict
${\cal B}^{\rm semi}_{\rm NR(S)}\sim$ a few $\times 10^{-4}$
and ${\cal B}^{\rm semi}_{\rm NR(S)}\simeq {\cal B}^{\rm semi}_{\rm NR(P)}$,
indicating that the non-resonant s-wave component alone cannot account for 
the total observed s-wave strength.

The $\kappa$ meson remains the most elusive of the light scalar states,
and its existence in weak decays has long been considered inconclusive.
However, the precise extraction of $f^+(0)$ allows for reliable predictions of the branching fractions
${\cal B}^{00,-+,{\rm semi}}_{\kappa}$, by treating the $\kappa$ as an explicit scalar contribution.
In particular, our result for ${\cal B}^{\rm semi}_{\kappa}$, which dominates the s-wave contribution,
suggests that a sizable $\kappa$ contribution has long been present in weak decays,
but has often been absorbed into the non-resonant background.

Our investigation provides new insight into the internal nature of the light scalar mesons.
In semileptonic decays $D\to M_1 M_2 \ell\nu_\ell$,
only the s-wave $M_1 M_2$ pair can undergo s-wave scattering,
potentially forming a molecular state~\cite{Branz:2007xp,Sekihara:2015iha,Hsiao:2023qtk}.
However, the small branching fraction ${\cal B}^{\rm semi}_{\rm NR(S)}=(3.8\pm 1.9)\times 10^{-4}$
indicates that such meson-meson scattering, and hence a $\kappa$-molecule formation,
is highly suppressed. In contrast, treating the $\kappa$ as a compact state
leads to a successful description of the data via the form factor $f^+(q^2)$.

If the $\bar \kappa^0$ is regarded as a compact state,
it may correspond to either a conventional $q\bar q$ meson~($|s\bar d\rangle$)
or a $q^2\bar q^2$ tetraquark ($|s\bar d u\bar u\rangle$)~\cite{Maiani:2004uc,Stone:2013eaa}.
Likewise, the light scalar mesons $S_n=|(u\bar u + d\bar d)/\sqrt{2}\rangle$ and
$S_{ns}=|(u\bar u + d\bar d)s\bar s/\sqrt{2}\rangle$
represent $q\bar q$ and $q^2\bar q^2$ configurations, respectively,
and mix to form physical states such as $f_0$, $\sigma_0$, and $a_0^0$.
Under $SU(3)$ flavor $[SU(3)_f]$ symmetry~\cite{Stone:2013eaa,
Cheng:2002ai,Wang:2009azc,Hsiao:2023qtk},
the corresponding $D^+ \to \bar\kappa^0$ form factors are related by
$F_{q\bar q}^{D^+\to\bar\kappa^0}=\sqrt{2}(f_K/f_\pi) F^{D^+\to S_n}$ and
$F_{q^2\bar q^2}^{D^+\to\bar\kappa^0} = \sqrt{2}(f_\pi/f_K) F^{D^+\to S_{ns}}$,
with $f_K/f_\pi = 1.2$ and $f_\pi/f_K = 0.8$ accounting for $SU(3)_f$ breaking~\cite{Gronau:1995hm,Wang:2023uea}.
Using $F^{D^+\to S_n} = 0.47\pm 0.02$ and
$F^{D^+\to S_{ns}} = 0.31\pm 0.02$ as inputs~\cite{Hsiao:2023qtk},
we obtain $F_{q\bar q}^{D^+\to\bar\kappa^0}=0.82 \pm 0.05$ and
$F_{q^2\bar q^2}^{D^+\to\bar\kappa^0}=0.36 \pm 0.02$.
The extracted value $f^+(0) = 0.39 \pm 0.08$ lies significantly below the $q\bar q$ expectation
and much closer to the tetraquark prediction, providing support
for a $q^2\bar q^2$ configuration of the $\kappa$ resonance.

In summary, we have performed a partial-wave analysis of the semileptonic decay
$D \to K\pi e \nu_e$, complemented by an investigation of the nonleptonic decays
$D \to K\pi\pi$ and $D \to K\pi\rho$. Using non-resonant $D \to K\pi$ form factors
$h(q^2)$ and $w_\pm(q^2)$, along with the $D \to \kappa$ form factor $f^+(q^2)$,
all extracted via a global fit, we determined the branching fractions:
${\cal B}^{\rm semi}_{\rm NR(S)}\equiv
{\cal B}_{\rm NR}(D^+\to (K^-\pi^+)_{\rm s-wave} e^+\nu_e)=(3.8\pm 1.9)\times 10^{-4}$,
${\cal B}^{\rm semi}_{\rm NR(P)}\equiv
{\cal B}_{\rm NR}(D^+\to (K^-\pi^+)_{\rm p-wave} e^+\nu_e)=(2.9\pm0.7)\times 10^{-4}$,
and ${\cal B}^{\rm semi}_{\kappa}\equiv
{\cal B}(D^+\to \bar\kappa^0 e^+\nu_e,\bar\kappa^0\to K^-\pi^+)=(2.8\pm 1.3)\times 10^{-3}$.
These results indicate that the $\kappa$ resonance can dominate the s-wave branching fraction,
overturning the long-standing interpretation in the literature, which assumed that
both ${\cal B}^{\rm semi}_{\kappa}$ and ${\cal B}^{\rm semi}_{\rm NR(P)}$ were negligible,
and that the s-wave strength originated mainly from the non-resonant component
${\cal B}^{\rm semi}_{\rm NR(S)}$.
The extracted form factor $f^+(0) = 0.39 \pm 0.08$ differs significantly
from the value expected for a conventional $q\bar{q}$ state ($0.82 \pm 0.05$),
and instead agrees with predictions for a compact tetraquark configuration ($0.36 \pm 0.02$).
This agreement provides additional support for interpreting light scalar mesons as tetraquark states.

\section*{ACKNOWLEDGMENTS}
This work was supported in part by the National Natural Science Foundation of China
(Grants No. 12575101 and No. 12175128).

\end{document}